\documentstyle[floats,prl,aps,epsf]{revtex} 
\begin{document}
\twocolumn[\hsize\textwidth\columnwidth\hsize\csname@twocolumnfalse\endcsname

\title{Computing the Complete Gravitational Wavetrain from Relativistic
	Binary Inspiral}

\author{Matthew D. Duez${}^1$, Thomas W. Baumgarte${}^1$, 
	and Stuart L. Shapiro${}^{1,2}$}

\address{
${}^1$ Department of Physics, University of Illinois 
 	at Urbana-Champaign, Urbana, IL 61801 \\
${}^2$ Department of Astronomy, \& NCSA, 
	University of Illinois at Urbana-Champaign, Urbana, IL 61801 }

\maketitle

\begin{abstract}
We present a new method for generating the nonlinear gravitational
wavetrain from the late inspiral (pre-coalescence) phase of a binary
neutron star system by means of a numerical evolution calculation in
full general relativity.  In a prototype calculation, we produce 214 wave
cycles from corotating polytropes, representing the final part of the
inspiral phase prior to reaching the ISCO.  Our method is based on the
inequality that the orbital decay timescale due to gravitational radiation
is much longer than an orbital period and the approximation that gravitational
radiation has little effect on the structure of the stars.  We employ
quasi-equilibrium sequences of binaries in circular orbit for the
matter source in our field evolution code.  We compute the 
gravity-wave energy flux, and, from this, the inspiral rate, at a discrete set
of binary separations.  From these data, we construct the
gravitational waveform as a continuous wavetrain.  Finally, we discuss
the limitations of our current calculation, planned improvements, and
potential applications of our method to other inspiral scenarios.
\end{abstract}

\draft

\vskip2pc]
%
%
%
%

The inspiral and coalescence of binary neutron stars are among the
most promising sources of gravitational radiation for the laser
interferometer gravitational wave observatories LIGO, VIRGO, GEO, and
TAMA, which should become operational in a few years.  It is estimated
\cite{motive} that LIGO/VIRGO will be able to detect the last 15
minutes of the inspiral (16,000 orbits) of a typical binary neutron
star system.  Detection and analysis of the signal will depend on
matched filter techniques, which require theoretical templates of the
waveforms.

The evolution of binary neutron stars proceeds in different stages.
By far the longest is the initial quasi-equilibrium {\it inspiral} phase,
during which the stars move in nearly circular orbits, while the
separation between the stars slowly decreases as energy is carried
away by gravitational radiation.  The quasi-circular orbits become
unstable at the innermost stable circular orbit (ISCO), where the
inspiral enters a {\it plunge and merger} phase.  The merger and
coalescence of the stars happens on a dynamical timescale, and
produces either a black hole or a larger neutron star, which may
collapse to a black hole at a later time.  The final stage of the
evolution is the {\it ringdown} phase, during which the merged object
settles down to equilibrium.

Two distinct approaches have commonly been employed in general 
relativity to study the
inspiral phase.  Much progress has been made in post-Newtonian (PN) studies
of compact binaries (see, e.g.~\cite{djs00}, and references therein).
Most of these approaches, however, approximate the stars as point
sources, which neglects important finite-size effects for neutron
stars.  Also, PN expansions may not converge sufficiently rapidly 
in the strong-field region near the ISCO.  
In an independent approach, relativistic quasi-equilibrium 
models of the inspiral phase of neutron star 
binaries have been constructed
numerically (see, e.g., \cite{matter} for corotating binaries
and~\cite{irrotational} for the more realistic case of irrotational
binaries).  Binaries have
also been evolved using fully relativistic hydrodynamics
\cite{Shibata_hydro,Shibata_hydro2}, but computational constraints
limit these evolutions to a few orbits.

In this paper, we propose a ``hydro without hydro'' method for using
the ``snapshots'' generated by quasi-equilibrium codes to create
a complete wavetrain~\cite{idea,HwoH}.  We use these snapshots to find the
gravitational waveform at a given separation and construct an
effective radiation reaction.  From this, the inspiral rate is known,
and the entire wavetrain can be constructed.  We describe this
method below.  We then present the results of a prototype calculation,
and discuss its implications.

%
%
%
%

The fundamental assumption in the quasi-equilibrium approximation is
that the orbital timescale is much shorter than the gravitational
radiation reaction time\-scale.  The binary inspiral then proceeds along
a sequence of quasi-equilibrium configurations in nearly circular
orbits with constant rest mass $M_0$.  This assumption significantly
simplifies the problem, since the hydrodynamical equations can be
integrated to yield a relativistic Bernoulli equation \cite{matter}.  
A similar approximation is routinely adopted 
in stellar evolution calculations.  There too, the evolutionary timescale
is much longer than the hydrodynamical timescale, so that the star can
safely be assumed to be in quasi-equilibrium on a dynamical timescale.

Constructing quasi-equilibrium binary configurations yields their total
mass-energy $M(r)$ and their orbital angular frequency 
$\Omega_{\rm orb}(r)$ as a
function of separation $r$.  The ISCO of an evolutionary $M_0 = const$
~sequence, where the assumption of quasi-equilibrium breaks down, can
be identified by locating a turning point along an $M$ versus $r$ curve
(see, e.g., the cross in Fig.~1).

Given a relativistic quasi-equilibrium configuration, we can compute
the emitted gravitational radiation by employing the binary data for the 
matter source terms in Einstein's field equations.  In
particular, the gravitational fields can be evolved in the presence of
these predetermined matter sources with a relativistic evolution code
(``hydro without hydro'',~\cite{HwoH}).  There is no need to re-solve
any of the hydrodynamic equations for the source, as we are assuming
that the stars move in circular orbits and are little affected by the
waves.  We thus simply rotate the source terms on our numerical grid.
Repeating this calculation for members of an evolutionary
sequence at discrete separations $r$, we obtain the wave amplitude
$A(r)$ and the orbit-averaged gravitational wave luminosity
$<dM/dt>(r)$ at these separations.  A fitting function can then be
used to interpolate to any intermediate separation (see Fig.~1).

The inspiral is determined by computing the effective radiation
reaction due to the radiation of mass-energy by gravitational waves.  
The energy loss $<dM/dt>$ forces the binary to a closer separation,
as it ``slides down'' the quasi-equilibrium $M$ vs $r$ curve.  
The inspiral rate is thus
\begin{equation}
\label{drdt}
{dr \over dt} = {<dM / dt> \over (dM / dr)_{\rm eq}},
\end{equation}
Solving this equation determines $r$ as a function of time
$t$, which then allows us to express $A(t)$ and $\Omega_{\rm orb}(t)$
in terms of $t$.  The complete quasi-equilibrium wavetrain can then be
constructed from
\begin{equation}
\label{h}
r_s h(t) = A(t)\sin\bigl(\int_0^t \Omega_{\rm GW}(t')dt' \bigr) \, ,
\end{equation}
where we expect $\Omega_{\rm GW} \approx 2\Omega_{\rm orb}$ due to the $m=2$
dominance of the radiation, and where $r_s$ is the distance between
the source and the distant observer.  The validity of our
method and its ability to produce reliable wavetrains is most easily
demonstrated in relativistic scalar gravitation, where the quasi-equilibrium
wavetrain can be compared to an exact numerical solution~\cite{ybs00}.

%
%
%
%

In the following we present a prototype calculation which employs the
quasi-equilibrium models of Baumgarte {\it et.al.}~\cite{matter}.
These models assume maximal slicing $K \equiv K^i_{~i} = 0$, where
$K_{ij}$ is the extrinsic curvature, and spatial conformal flatness, so
that the spatial metric $\gamma_{ij}$ can be written as a product of
the conformal factor $\psi$ and a flat metric $f_{ij}$, $\gamma_{ij} =
\psi^4 f_{ij}$.  The latter is believed to approximately minimize the
gravitational wave content of the spacetimes \cite{York} while 
introducing only small deviations from correct evolutionary 
sequences \cite{test_conformal}.  The stars are modeled
as corotating polytropes $P = \kappa\rho_0^{1+1/n}$ with polytropic
index $n=1$.  We non-dimensionalize our results by setting $\kappa =
1$.  In addition to the matter profile, these models provide the
conformal factor $\psi$, the lapse function $\alpha$ and the shift
vector $\beta^i$.

We insert these quasi-equilibrium data into our relativistic evolution
code, which is based on a conformal decomposition of the ADM
equation~\cite{code}.  This decomposition conveniently splits the
fundamental variables into longitudinal quantities ($\psi$ and $K$)
and ``wave-like'' transverse quantities (the conformally related
metric $\tilde \gamma_{ij}$ and the tracefree part of the extrinsic
curvature $\tilde A_{ij}$).  We also introduce ``conformal connection
functions'' $\tilde \Gamma^i$ as independent auxiliary quantities,
which improves numerical stability dramatically (cf.~\cite{nok87}).

We insert ``by hand'' the matter variables, the gauge variables
($\alpha$ and $\beta^i$) and the longitudinal quantities ($\psi$ and
$K = 0$) for all times, appropriately rotated by $d\phi/dt =
\Omega_{\rm orb}$~\cite{footnote1}.  The transverse variables ($\tilde
\gamma_{ij}$, $\tilde A_{ij}$ and $\tilde \Gamma^i$) are evolved
dynamically, adopting the quasi-equilibrium data $\tilde A_{ij}$,
$\tilde \gamma_{ij} = f_{ij}$ and $\tilde \Gamma^i = 0$ as initial
data.  We found that we could reduce the numerical errors by explicitly
splitting quasi-equilibrium ``background'' terms from dynamically evolved
terms in the evolution equation for the $\tilde \Gamma^i$.

\begin{figure}
\epsfxsize=3.0in
\begin{center}
\leavevmode
\epsffile{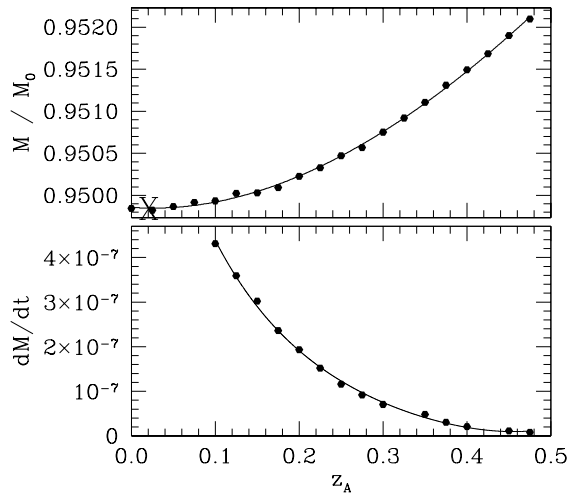}
\end{center}
\caption{Total mass $M$ and gravitational wave luminosity $dM/dt$
	versus separation for our evolutionary sequence.  Both 
	quantities are measured by a distant observer.  
	The separation is parameterized by $z_A$ (see text).  The points
	represent individual numerical models, and the solid lines are
	fits.  The turning point $dM/dz_A = 0$ in the top panel locates
	the ISCO (marked by the cross). }
\end{figure}

We evolve on a 120$\times$120$\times$60 zone grid, imposing equatorial
symmetry across the z-axis.  Each star is covered by roughly 8 points
across a diameter.  We impose approximate outgoing wave boundary
conditions on the transverse variables at a separation of
$0.1\lambda_{\rm GW}$ - $0.3\lambda_{\rm GW}$, where $\lambda_{\rm
GW}$ is a gravitational wavelength.  We expect our outer boundary 
location to be one of the
primary sources of error (see discussion below, and compare
with~\cite{Shibata_hydro2} where the boundaries are similarly placed).  
We monitor the violation in the Hamiltonian and momentum
constraints and find that they are all satisfied quite well and that
the degrees of violation do not grow significantly with time.  We
doubled the resolution for one configuration, and found that $A$
changed by only $\approx$ 2\%.

\begin{figure}[t]
\epsfxsize=3.0in
\begin{center}
\leavevmode
\epsffile{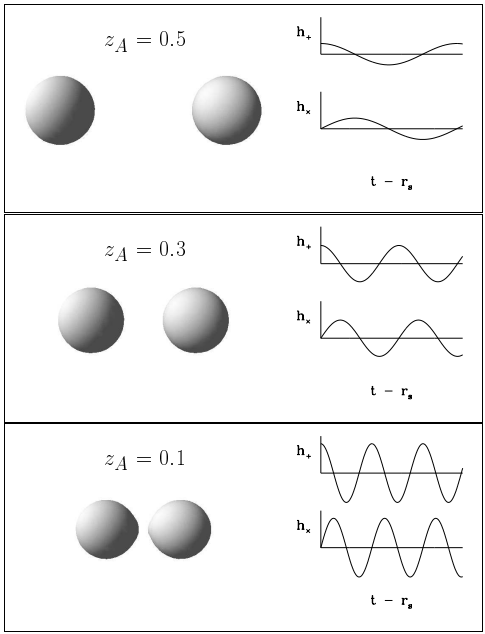}
\end{center}
\caption{Three representative binary configurations with their
	corresponding waveforms.  We show the binaries with greatest
	and smallest separation in our sequence, and one intermediate 
	member.
	At right, the two polarizations of gravity waves ($h_+$ and
	$h_{\times}$) along the $z$-axis are shown.  On the $z$-axis,
	the two polarizations only differ by a phase.}
\end{figure}

To extract gravitational waves, we match to even-parity Moncrief
variables $\psi_{lm}$ (or Zerilli functions) at the outer part of the
grid (see \cite{Moncrief,Perturb_Match,Shibata_waves} for details).
For simplicity, we focus on $l=2$, $m=\pm 2$ modes, which we expect to be
dominant.  We have checked that the contribution to the energy flux of
the $l=4$, $m=\pm 4$ mode is very small ($\sim 4$\%), and that of the
$l=2$, $m=0$ mode negligible ($< 10^{-6}$).  Assuming $l=2$, $m=\pm 2$
dominance, we have
\begin{eqnarray}
\displaystyle
\frac{dM}{dt} & = & \displaystyle 	
	\frac{1}{32\pi} \left( (\partial_t \psi_{22})^2
	+ (\partial_t \psi_{2-2})^2 \right) \\
A & = & \displaystyle \sqrt{ {5\over 16\pi}\left(\overline\psi^2_{22} 
			+ \overline\psi^2_{2 -2} \right) }
\end{eqnarray}
where $\overline\psi_{2\pm 2}$ denotes the amplitude of $\psi_{2\pm
2}$.  We approximate the asymptotic value of $\overline\psi_{lm}$ by
its value at the edge of our numerical grid.  Due to our neglect of
self-consistent transverse fields in the initial data, we find a
transient noise in the waveforms during the first light-crossing time.
After that time we observe the expected periodic waveforms originating
from the periodic binary orbit.


\begin{figure}[t]
\epsfxsize=3.0in
\begin{center}
\leavevmode
\epsffile{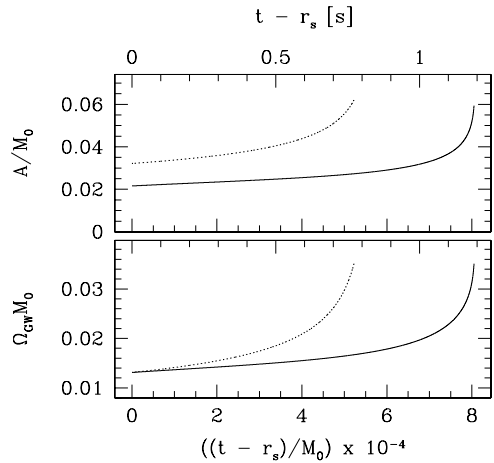}
\end{center}
\caption{The amplitude $A$ and angular frequency $\Omega_{\rm GW}$ of
	the gravitational wave signal as a function of retarded time
	$t-r_s$ (solid line).  The dotted line is the result for
	Newtonian point masses in the quadrupole approximation.  The
	top label provides times in seconds for a binary of total
	rest mass $M_0 = 2 \times 1.5 M_{\odot}$.  Our sequence ends 
	at a frequency $\Omega_{\rm GW} = 2.36$ kHz for such a binary.}
\end{figure}

We constuct 14 different configurations along a sequence of constant
total rest mass $M_0 = 2 \times 0.1$ in our nondimensional units.
The rest mass of the individual stars is about 55\% of the maximum
allowed rest mass of an isolated, nonrotating star.  We parameterize
the binary separation by the dimensionless quantity $z_A$, the ratio
of the separation between the innermost points and the outermost
points on the stars (see~\cite{matter}).  In terms of $z_A$, touching
binaries correspond to $z_A = 0$ and infinitely separated
configurations to $z_A = 1$.  In the top panel of Fig.~1, we plot the
ADM mass $M$ as a function of separation, and in the bottom panel the
gravitational wave luminosity $dM/dt$ as inferred from our ``hydro
without hydro'' integrations.  Since we expect our quasi-equilibrium
method to break down at the ISCO, we construct waveforms for
configurations with $z_A \geq 0.1$, which is close to but safely
outside the ISCO.  In Fig.~2 we illustrate three of these
configurations together with their extracted waveforms.  In Fig.~3 we
show the wave amplitude $A$ and the gravitational wave frequency
$\Omega_{\rm GW}$ as a function of retarded time.  The solid line is
our numerical result, which we compare with the wavetrain of a
Newtonian point mass system as computed from the quadrupole
approximation (dotted line).  The difference is caused by differences
in the binding energy and the gravitational wave luminosity due to 
relativity, as well
as the numerical limitations of our calculation (see discussion
below).  Finally, we construct the continuous gravitational wavetrain
from $A$ and $\Omega_{\rm GW}$, shown in Fig.~4.

\begin{figure}[t]
\epsfxsize=3.0in
\begin{center}
\leavevmode
\epsffile{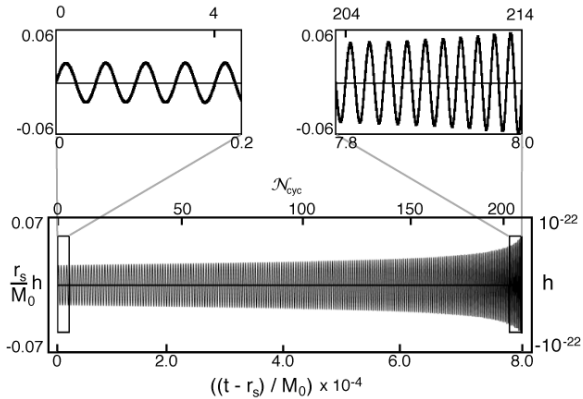}
\end{center}
\caption{The full wavetrain $h_+$ or $h_{\times}$ on the $z$-axis as
	a function of retarded time and cycle number.  The top two
	panels show enlargements of the first four and last ten
	cycles.  We also indicate the gravitational strain $h$ for a
	binary of total rest mass 
	$M_0 = 2 \times 1.5 M_{\odot}$ at a separation of 100 Mpc.}
\end{figure}


In summary, we propose a new technique to construct gravitational
wavetrains from inspiraling relativistic binaries and demonstrate
the feasability of this technique by a numerical example.  Our scheme is 
extremely efficient computationally.  For the computational cost 
of 14 orbits, we produced the complete wavetrain covered by our 
sequence (214 cycles).  There is no reason why one could not, with more points 
along the sequence, produce many more cycles of the inspiral.   

It is clear, however, that several improvements must be implemented to
produce more accurate and reliable wavetrains.  Our outer boundary
conditions are crude and imposed at too small a separation.  To gauge
the error of our simplistic wave extraction, we measured
$\overline\psi_{22}$ at 0.23$\lambda_{\rm GW}$ and at
0.35$\lambda_{\rm GW}$ and found a difference of roughly 10\%.
Gravitational waves should be extracted in the wave zone $r_s >
\lambda_{\rm GW}$, or more sophisticated boundary conditions and wave
extraction should be applied (e.g.~\cite{outer_BC}).  Furthermore, the
assumption of corotation is not realistic; using irrotational
sequences is more physical \cite{irr_1}.  Finally, one could consider
iterating this method by adding the radial velocity to the source and
incorporating the calculated transverse components of the field, as
well as deviations from conformal flatness, and then recomputing the
waves and inspiral.

The binary black hole inspiral problem could be approached in a 
similar way \cite{bek}.  This is a prospect we are currently investigating.



It is a pleasure to thank Masaru Shibata for stimulating discussions
and useful suggestions.  We would also like to thank REU students
H. Agarwal, E. Engelhard, K. Huffenberger, P. McGrath, J. Mehl, and
D. Webber for assistance on Figures 2 and 4.  TWB gratefully acknowledges
support through a Fortner Fellowship.  Much of the calculation
and visualization were performed at the National Center for
Supercomputing Applications at UIUC.  This paper was supported in part
by NSF Grants AST 96-18524 and PHY 99-02833 and NASA Grant NAG 5-7152
to UIUC.


\begin{references}
\bibitem{footnote1} We have verfied that dynamically evolving $\psi$,
	$K$ and solving maximal slicing to find $\alpha$ at each timestep
	yields very similar results.
\end{references}
\end{document}